\documentclass[sigconf]{acmart}

\AtBeginDocument{%
  \providecommand\BibTeX{{%
    \normalfont B\kern-0.5em{\scshape i\kern-0.25em b}\kern-0.8em\TeX}}}

\setcopyright{none}
\copyrightyear{.}
\acmYear{.}
\acmDOI{.}

\acmConference{.}
\acmBooktitle{.}
\acmPrice{.}
\acmISBN{.}
\renewcommand\footnotetextcopyrightpermission[1]{} % removes footnote with conference information in first column
\usepackage{tikz}
\usepackage{amsmath}
\usepackage[english]{babel}

\usepackage{booktabs}
\usepackage{xurl}
\usepackage{tabularx}
\usepackage{xcolor}
\usepackage[binary-units]{siunitx}
\usepackage{setspace}
\usepackage{acronym}
\usepackage{rotating}
\usepackage{subcaption}
\usepackage{units}
\usepackage{siunitx}
\usepackage{multirow}
\usepackage{listings}
\usepackage{pgfplots}
\usepackage{pgfplotstable}
\usepgfplotslibrary{external}
\usepackage[shortlabels]{enumitem} 
\microtypecontext{spacing=nonfrench}
\usepackage{mathtools}
\usepackage{tikz-cd}
\usepackage{graphicx}
\usepackage{xspace}
\usepackage{pgf-pie}
\usepackage{lipsum}

\newacro{VM}{virtual machine}
\newacro{MPU}{memory protection unit}
\newacro{FaaS}{Function as a Service}

\pgfplotsset{compat=1.17}

\newcommand{\cfc}{{CertFC\xspace}}% can be replaced by {CertFemto-container} or {CertrBPF}
\newcommand{\dx}{\ensuremath{\partial x}\xspace}

\newcommand{\shepherd}[1]{#1} 

\newcommand\blfootnote[1]{%
  \begingroup
  \renewcommand\thefootnote{}\footnote{#1}%
  \addtocounter{footnote}{-1}%
  \endgroup
}

\begin{document}

%\title{Femto-Containers: Ultra-Lightweight Software Module Virtualization \& Isolation for IoT Microcontrollers}
%\title{Femto-Containers: DevOps on Microcontrollers with Lightweight Virtualization \& Isolation for IoT Software Modules}

\title[Femto-Containers]{
%\framebox[1.01\width]{\parbox{\dimexpr\linewidth-2\fboxsep-2\fboxrule}{If you cite this paper, please use the ACM MIDDLEWARE'22 reference: K. Zandberg, E. Baccelli, S. Yuan, F. Besson, JP Talpin.  Femto-Containers: Lightweight Virtualization and Fault Isolation For Small Software Functions on Low-Power IoT Microcontrollers. In Proc. of 23rd ACM/IFIP MIDDLEWARE, Nov. 2022.}}
Femto-Containers: Lightweight Virtualization \\ and Fault Isolation For Small Software Functions \\ on Low-Power IoT Microcontrollers
}


\author{Koen Zandberg}
\affiliation{%
  \institution{Inria}
%  \city{City}
  \country{France}
}

\author{Emmanuel Baccelli}
\affiliation{%
  \institution{Inria}
%  \city{City}
  \country{France}
}
\affiliation{%
  \institution{Freie Universit\"at Berlin}
%  \city{City}
  \country{Germany}
}

\author{Shenghao Yuan}
\affiliation{%
  \institution{Inria}
%  \city{City}
  \country{France}
}

\author{Frédéric Besson}
\affiliation{%
  \institution{Inria}
%  \city{City}
  \country{France}
}

\author{Jean-Pierre Talpin}
\affiliation{%
  \institution{Inria}
%  \city{City}
  \country{France}
}

\renewcommand{\shortauthors}{.}

\begin{abstract}
\blfootnote{If you cite this paper, please use the ACM MIDDLEWARE'22 reference: K. Zandberg, E. Baccelli, S. Yuan, F. Besson, JP Talpin.  Femto-Containers: Lightweight Virtualization and Fault Isolation For Small Software Functions on Low-Power IoT Microcontrollers. In Proc. of 23rd ACM/IFIP MIDDLEWARE, Nov. 2022.}
Low-power operating system runtimes used on IoT microcontrollers typically provide rudimentary APIs, basic connectivity and, sometimes, a (secure) firmware update mechanism.
In contrast, on less constrained hardware, networked software has entered the age of serverless, microservices and agility.
With a view to bridge this gap, in the paper we design Femto-Containers, a new middleware runtime which can be embedded on heterogeneous low-power IoT devices. 
Femto-Containers enable the secure deployment, execution and isolation of small virtual software functions on low-power IoT devices, over the network. % in a fashion resembling Function-as-a-Service (FaaS) adapted to IoT use cases.
We implement Femto-Containers, and provide integration in RIOT, a popular open source IoT operating system. 
\shepherd{We then evaluate the performance of our implementation, which was formally verified for fault-isolation, guaranteeing that RIOT is shielded from logic loaded and executed in a Femto-Container.
%Finally, we evaluate experimentally the performance of Femto-Containers in a variety of use cases, 
Our experiments on various popular microcontroller architectures (Arm Cortex-M, ESP32 and RISC-V) show that Femto-Containers offer an attractive trade-off in terms of memory footprint overhead, energy consumption, and security.}

\end{abstract}

\iffalse



\fi

\iffalse
\begin{teaserfigure}
  \includegraphics[width=\textwidth]{sampleteaser}
  \caption{Seattle Mariners at Spring Training, 2010.}
  \Description{Enjoying the baseball game from the third-base
  seats. Ichiro Suzuki preparing to bat.}
  \label{fig:teaser}
\end{teaserfigure}
\fi

\maketitle

\section{Introduction}

An estimated 250 billion microcontrollers are in use today~\cite{250billionMCU}.
An increasing percentage of these microcontrollers are networked and take part in distributed cyber-physical systems and the Internet of Things (IoT) we increasingly depend upon.
\shepherd{For example, such low-power microcontrollers are at the core of hundreds of millions of connected machines such as sensors and actuators relied upon not only in smart homes, but also in other networked areas of the IoT and industrial contexts (Industry 4.0).
On such hardware, the total memory available (for the whole system) is in the order of tens or hundreds of kBytes, without virtual memory management (MMU), often also without hardware memory protection (MPU).
Neither Linux (or derivatives/equivalents) nor traditional hypervisor can be used as software platform on such hardware, and in effect the challenge of deploying and maintaining distributed low-power IoT software is exacerbated.}

Recently, with the wider availability of low-power operating systems alternatives~\cite{hahm2015operating} and adequate network stacks, low-power IoT software has made giant leaps forward; but fundamental gaps remain compared to current practices for networked software.
In fact, current state-of-the-art for managing, programming, and maintaining fleets of low-power IoT devices resembles more PC system software workflow from the 1990s than today's common software practices. Simplistic application programming interfaces (APIs) offer basic performance and connectivity, but no additional comfort.

However, since the 1990s, networked software was revolutionized many times over.
Networked software has entered the age of server-less, micro-services and agility.
In the field, software modules that are deployed and running are expected to be quickly updatable in terms of functionalities, and in terms of bug fixes.
Additional layers and primitives providing cybersecurity, flexibility and scalability became crucial: virtual machines, script programming (\emph{e.g.} Python, Javascript), lightweight software containerization (\emph{e.g.} Docker, Function-as-a-Service~\cite{fox2017status} etc.). DevOps~\cite{bass2015devops} workflow drastically shortened software development/deployment life cycles to provide continuous delivery of higher software quality.

%hypervisors and software virtualization, deployment and management tools for swarms of virtualized software instances (e.g., Kubernetes or AWS), and frameworks for decentralizing system software updates, development and maintenance on platforms such as Linux, Android, iOS, Windows etc.

In such a context, low-power IoT devices based on microcontrollers are the new ‘weakest link’ within distributed cyber-physical systems.
%State-of-the-art scripting laguages for low-power IoT devices (microPython, JerryScript, Arduino) typically not applicable beyond rapid prototyping, as they do not exploit low-power modes of the microcontrollers.
Indeed, state-of-the-art primitives for 'serverless', including lightweight virtual machine (VM) runtimes~\cite{morabito2018lightweight-vm} and container runtimes (\emph{e.g.} Docker) are not applicable on low-power devices: they typically depend on operating systems and larger resources which don't fit microcontrollers. In particular, VMs are either too prohibitive in terms of hosting engine memory resource requirements (\emph{e.g.} standard Java virtual machines), or restricted to very specific use cases (\emph{e.g.} JavaCard).
This lackluster creates bottlenecks that severely impact both flexibility and cybersecurity in the low-power IoT space.

%A key question emerges: can we provide adequate new concepts for ultra-lightweight virtualization, software containerization and rapid deployment on swarms of IoT devices, combining agility, low-power consumption and cybersecurity?
{\bf Goals of this paper --} We aim to design a middleware function runtime adequate for heterogeneous microcontrollers. 
Mainly enabling the secure deployment, execution and isolation of small virtual software functions on low-power IoT devices. % using software virtualization
%, providing fault isolation, which are adequate for heterogeneous low-power IoT devices based on microcontrollers. %Several functions could run in parallel on the same microcontroller. 
What we aim for in priority is small start-up time for deployed functions, and negligible overhead w.r.t. memory footprint on the microcontroller when adding a hosted function runtime to the OS, compared to the same functionality implemented natively (in the OS).

%In this paper, we do not focus on DevOps backend aspects, which is not the bottleneck in our view. Instead, we focus on the embedded software architecture aspects which enable small software module hosting and isolation on IoT microcontrollers. We also consider protocols aspects which heterogeneous low-power IoT can interconnect with potential DevOps backends, for securing software updates over the network.

{\bf Contributions --}
In this paper, the work we present mainly consists in the following:
\begin{itemize}
%\item we survey existing techniques for process isolation \& virtualization for microcontrollers;
%\item We design Femto-Containers, a novel approach for the virtualisation and isolation of multiple software functions running on constrained IoT devices;
\item \shepherd{We propose Femto-Containers, a novel middleware for the abstraction, secure deployment, execution and isolation of (multiple, concurrent) software functions on heterogeneous microcontroller-based IoT devices. We design Femto-Containers as an extension of rBPF virtual machines}; 
%We design Femto-Containers around an existing virtual machine implementation as extension optimized for a Function as a Service-like architecture.
%which can host and isolate multiple run-time containers on a low-power microcontroller, with very small memory footprint overhead;
%\item We compare different ultra-lightweight VMs: MicroPython, WebAssembly (WAMR), JavaScript (RIOTjs) and eBPF. We show that comparatively, a Femto-Container runtime as extension of eBPF requires up to 10x less memory footprint
\item We benchmark ultra-lightweight virtualization techniques based on Python, WebAssembly, JavaScript and eBPF. We show that, comparatively, a Femto-Container runtime based on eBPF virtualization requires 10x less memory footprint;
\item \shepherd{We provide an open source implementation of Femto-Containers, which we integrate in practice on a common, general-purpose operating system for low-power networked microcontrollers (RIOT)};
\item \shepherd{We formally verify key components of our Femto-Container implementation, guaranteeing fault-isolation amongst concurrent Femto-Containers, and the underlying OS};
\item We evaluate the performance of Femto-Containers in a variety of use cases, on the most popular 32-bit microcontroller architectures: Arm Cortex-M, ESP32 and RISC-V. %We show Femto-Containers offer an attractive trade-off in terms of memory footprint overhead, energy consumption, and security.

%\item \todoEB{??? We design and implement primitives enabling the secure update of Femto-containers over arbitrary networks, which we demonstrate and evaluate over several low-power IoT transport (BLE and IEEE 802.15.4 radios) ???;}
\end{itemize}
%\section{Analysis}
%\section{Scenarios \& Threat Model}

% Needs an intro sentence
%\section{FaaS-like Scenarios on Microcontrollers}
\section{Multi-Tenant Software Scenarios on Microcontrollers}
\label{sec:case-study}
Software on low-power IoT devices is growing complexity, driven by cybersecurity, interoperability, and device management requirements. 
%of , maintenance and update mechanisms also grow in complexity.
In practice, the development of embedded software components is therefore often delegated to distinct entities.
For security and privacy reasons these distinct entities have limited mutual trust~\cite{thomas2018design}.
% As low-power embedded IoT software complexifies on various devices, it becomes necessary for security (and sometimes also privacy) reasons to delegate maintenance and updates of different parts of the embedded software to distinct entities with limited mutual trust (as described in~\cite{thomas2018design} for instance).
\shepherd{For example, in this context, prior work such as Amulet~\cite{hardin2018mem-isolation-MCUs} aim to isolate multiple applications from each other on a microcontroller, and to protect the underlying OS from application code.}
Furthermore, maintenance of these low-power IoT devices typically requires on-the-fly instrumentation.
Safety requires that hot insertion of instrumentation code cannot break running software (already deployed).
\shepherd{For example, prior work such as TockOS~\cite{tockos} aims to enable safe multiprogramming of low-power microcontrollers.}

\begin{figure}[t]
  \centering
\includegraphics[width=\columnwidth]{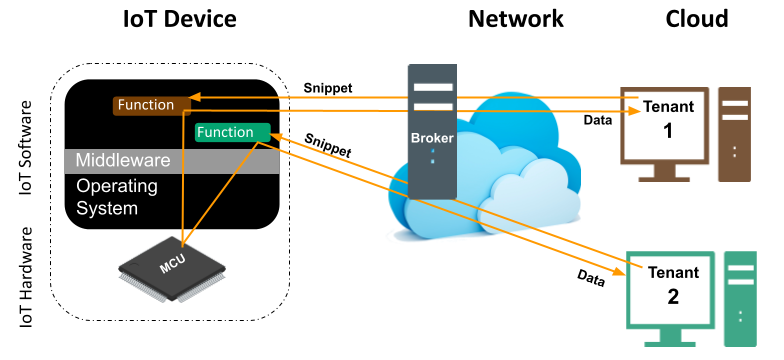}
  \caption{Container runtime use on IoT microcontrollers.}
  \label{fig:femto-containers-use-cases}
\end{figure}

In this context, we identify the following categories of use-cases, depicted in \autoref{fig:femto-containers-use-cases}:
\begin{enumerate}
  \item Use-case 1: Hosting and isolation of a high-level business function.
        This function can be updated securely, on-demand, remotely over the low-power network.
        The execution of this type of logic is typically periodic in nature, and has loose (non-real-time) timing requirements.
  \item Use-case 2: Hosting and isolation of debug and monitoring code functions at low-level.
        These are inserted and removed on-demand over the network.
        The functions must not interfere with existing code on the device.
        Comparatively, this type of function is short-lived and exhibits stricter timing requirements.
  \item Use-case 3: Hosting and isolation of several functions, managed by several different tenants.
\end{enumerate}

%Users in the above scenarios are provided with an event-driven programming model whereby %applications are scheduled and distributed by a management broker over fleets of %interchangeable
%workers/machines.
Users in the above scenarios are provided with an event-driven programming model and fine-grained computational space, hosted on-demand on fleets of designated low-power IoT devices. Furthermore, the API available for a function fundamentally abstracts away most of the hardware and the OS.
Conceptually, these scenarios thus partly mimic a Function-as-a-Service (FaaS) programming model~\cite{fox2017status}. 
%FaaS is a current trend amongst large cloud providers, which aims to provide customers with fine-grained computational space on-demand, paid for on a per-execution basis.
%Compared to an approach based on containers (\emph{e.g.} Docker), FaaS provides an additional abstraction layer for both the operating system and the storage database: developers can thus purely focus on functionality.

%\paragraph{\bf Similarities with FaaS}  As with FaaS, users in our scenario are provided with an event-driven programming model whereby applications are scheduled and distributed by a management broker over fleets of %interchangeable
%workers/machines, which ensure both automatic scaling and high availability, via the network.
%
%Similarly to FaaS, users in our scenario are provided with fine-grained computational space, on-demand, hosted on designated low-power IoT devices. Furthermore, the API available for a function fundamentally abstracts away most of the hardware and the OS, as in FaaS.

As with FaaS, \shepherd{multiple functions provided by distinct stakeholders must run on a single device. For example OEM firmware may have to be completed/customised by separate components, e.g. different developers/tenants may provide drivers separately, which should be fault-isolated and restricted to using only driver-relevant resources. Meanwhile, OS maintainers can deploy/run debug snippets inserted elsewhere in the embedded code.}

However, as a stable high-throughput network connection cannot be assumed, storage capability is restricted to device-local storage.
%koen: Mention that we run a large number of container across a large number of devices instead of a large number of containers on a single large server (as with FaaS)
Furthermore, \shepherd{scalability aspects of FaaS and container techniques (e.g. running thousands of containers on a single machine) do not play a significant role here. Instead, the scenarios we identify above require running just a handful of isolated functions on top of the embedded OS. However, considering potentially large fleets of IoT devices, the scenario may nevertheless involve a large number of containers (but across a large number of devices).}
%\paragraph{\bf Differences with FaaS} 

% Koen: This should be emphasised in the introduction of the paper, and summarized here. I would refrain to refering back to buzzwords such as industry 4.0 here again.
In this paper, we focus primarily on the middleware embedded in the devices: the runtime permitting to host, run and isolate functions, on-demand, on heterogeneous low-power IoT devices deployed in the field, \shepherd{based on popular 32-bit microcontroller architectures (such as the ARM Cortex-M class, RISC-V or ESP32).}
%Such microcontrollers are at the core of hundreds of millions of connected machines such as sensors and actuators relied upon not only in smart homes, but also in other networked areas of the Internet of Things and industrial contexts (Industry 4.0). On such hardware, the total memory available (for the whole system) is in the order of tens or hundreds of kBytes, without virtual memory management (MMU), often also without hardware memory protection (MPU). Neither Linux (or derivatives/equivalents) nor traditional hypervisor or containers such as Docker can be used on such hardware, and in effect the challenge is exacerbated.
%Therefore, the management broker aspects of FaaS are mostly ignored here. Moreover, FaaS typically manages the automatic scaling/distribution of a function over a set of interchangeable devices. In our scenario, this is not desirable: on the contrary, devices must be selected based on their cyberphysical properties (specific sensors/actuators, geographical/topological location etc.).

\section{Threat Model}
\label{sec:isolationmodel}
%\paragraph{Threat Model.} 
When a client deploys functions on a device operational in the field, the embedded environment has to ensure these functions are sandboxed.
%that any function running is isolated. 
In our threat model, we consider both malicious tenants which can deploy malicious code and malicious clients which can maliciously interact with deployed code~\cite{ren2020fine}.
%Both the malicious client and the malicious tenant aim to compromise the system and abuse resources and information contained on the host device.
%A tenant supplies the application for the virtual machine. The tenant is able to upload potentially untrusted application to the system for the virtual machine. Second is the client interacting with the provided application. The client can send requests to networked virtual machine applications of different tenants. These two parties have different mechanisms to potentially influence the host system or other tenants on the system.

\emph{\bf Malicious Tenant}:
The malicious tenant seeks to gain elevated permissions on the device it has already a set of permissions on.
This tenant is already allowed to run code in the sandboxed environment, and the tenant might want to break free from the sandbox to either the host system or a different sandbox it doesn't have permissions for.
While a tenant has to work within the permissions granted by the host service, it can make free use of the granted resources.
%A tenant could provide a vulnerable or malicious application code for execution inside a function.
%To protect against this, the host system must ensure that hosted functions are constrained in the resources they can access and have a fair share of the processing time and network bandwidth. % Maybe already too much of a solution.

\emph{\bf Malicious Client}:
The malicious client doesn't have any permissions for running sandboxed code on the device.
The only access the malicious client has is access to networked endpoints exposed by the device, e.g. CoAP endpoints exposed by existing sandboxed environments. The malicious client seeks to gain any permission on the device to influence it or gain access to confidential data on the device.
The malicious client could make use of an already vulnerable tenant function.
%The client can send requests to networked application, including arbitrary packets.
%Assuming a vulnerable application, the client could access any resource accessible by the tenant function.
%Protection of already vulnerable applications is considered out of scope.
%However, the hosting engine must isolate other tenants' memory access from a compromised tenant function.

%{\bf Online vs Offline Isolation Guarantees.} 

A number of attack vectors are considered in this work:
\begin{itemize}
\item \emph{Install and update time attacks}: These attacks focus on modifying the application during the transport to the sandbox environment. This includes man-in-the-middle modifications to the applications.
\item \emph{Privilege escalation to a different sandbox}: This class of attacks focus on escaping the sandbox of the application to a different sandbox. The new sandbox could have different permissions.
\item \emph{Privilege escalation to the operating system}: This attack class attempts to escape the sandboxed environment altogether to the operating system.
\item \emph{Resource exhaustion attacks}: The devices considered here have very limited resources, both computational power and battery energy are limited. A denial of service vector can be to exhaust these resources.
\end{itemize}

Within the Femto-Container design we consider network attacks to exhaust resources on the system to be out of scope, this type of attack should be guarded against by the embedded operating system.

\section{Related Work}

%\section{Related work on Lightweight VM Runtimes}

Typically, the fundamental building block for middleware embedded on devices allowing for generic function deployment and execution is a \emph{virtual machine runtime}. 

The vast majority of prior work on lightweight virtualization runtimes~\cite{morabito2018lightweight-vm} does not target microcontrollers, but microprocessor-class computers. Recent examples include for instance AWS Firecracker~\cite{firecracker-2020} for serverless computing, WebAssembly~\cite{haas2017wasm} for process isolation in Web browsers, or eBPF~\cite{mccanne1993bpf,fleming2017eBPF} for debug and inspection code inserted in the Linux kernel at run-time.

However, some ultra-lightweight virtualization approaches have been proposed for microcontrollers.
For example, minimized WebAssembly runtimes adapted to run on 32-bit microcontrollers were proposed, such as WAMR~\cite{wasm-micro-runtime} and WASM3~\cite{wasm3}.
RapidPatch~\cite{rapidpatch2022} uses an eBPF runtime to provide a hotpatching framework for RTOS firmwares.%, and the rBPF~\cite{zandberg2020minimal} provided a port of the eBPF instruction set to microcontrollers.

VM runtimes for microcontrollers include also earlier examples such as Mate~\cite{levis2002mate} or Darjeeling~\cite{darjeeling}, a subset of the Java VM, modified to use a 16 bit architecture, designed for 8- and 16-bit microcontrollers.
JavaCard~\cite{identity2018javacard} also uses a small Java virtual machine % \ac{VM}
tailored for cryptographic purposes, running on smart cards.

Recently, tiny scripted logic interpreters and runtimes have also been proposed to provide a basic virtualization environment. For instance, MicroPython~\cite{micropython} is a very popular scripted logic interpreter used on microcontrollers.
%Another popular scripted logic interpreter is JerryScript, which offers full ECMA5.1 scripting support.
Small Python runtimes are used on ESP8266 microcontrollers in prior work such as NanoLambda~\cite{george2020nanolambda}.
Small Javascript runtimes are used on Cortex-M microcontrollers in prior work such as RIOTjs~\cite{riot-js-container}.
However, complementary mechanisms should however be used to guarantee mutual isolation between scripts (such as SecureJS~\cite{ko2021securejs}).

\shepherd{The most closely related work was published in \cite{zandberg2020minimal} and in \cite{cavpaperanonymized}. In \cite{zandberg2020minimal}, authors provide rBPF, a port  of the eBPF instruction set in order to host a (single) VM on a microcontroller. In contrast, we extend rBPF's instruction set architecture and VM core with an adequate embedded loading and execution environment, which caters for well-defined, event-driven, short lived and isolated (concurrent) execution of (multiple) functions to be deployed on-the-fly on a networked microcontroller. On the other hand, while \cite{cavpaperanonymized}, concerns formal verification of the sole rBPF instruction interpreter, here we also verify the pre-flight instruction checker and we integrate both in our femto-container implementation -- the performance of which we thoroughly evaluate on various microcontrollers.
Our design and implementation are so small (a few hundreds lines of code) that formal verification was indeed realistic. Comparatively, software alternatives (WASM, MicroPython...) would require hundreds of thousands of lines of code, and magnitudes more lines of proof, all but voiding concrete perspectives of formal verification. Hardware alternatives (e.g. TrustZone~\cite{pinto2019trustzone}) are, to the best of our knowledge, not formally verified, and may require a software API to avoid unspecified hardware behaviours, hence faults.}

To the best of our knowledge, our work provides the first formally verified middleware based on eBPF virtualization able to host multiple tiny runtime containers on a wide variety of heterogeneous low-power microcontrollers.

\iffalse
We have already provided a survey of related work in~\autoref{sec:techniques-survey}.
To the best of our knowledge, the closest related work is rBPF~\cite{zandberg2020minimal}. Compared to rBPF:
\begin{itemize}
\item Femto-Containers are applicable not only to single container use-cases, but also to use cases hosting multiple containers and distinct tenants, concurrently, on the same microcontroller;
\item Femto-Containers provide additional containerization, security and isolation mechanisms;
\item Femto-Containers improve performance in the single-container case.
\end{itemize}
Furthermore, on the experimental side, we compare the performance of more diverse containers techniques, on a wider variety of low-power IoT hardware architectures.
\fi

\section{Embedded Runtime Architecture Design}
\label{sec:design}
\label{sec:femto-arch-design}

In this section, we introduce Femto-Containers, a new embedded runtime architecture tailored for constrained IoT devices, as described in the following.

Similarly to a FaaS runtime, Femto-Containers allow for the we deployment and execution of small logic modules.
%as shown in \autoref{fig:arch-design}.1. 
These modules, or functions, are hosted on top of a middleware offering isolation, abstraction and tight isolation with respect to the underlying OS and hardware. 
By combining isolation and hardware/OS abstraction, we retain the crucial properties of FaaS runtimes: code mobility and cyber-security. Differently from typical FaaS runtimes, however, Femto-Containers must be able to interact with specific hardware (\emph{e.g.} sensor/actuators), and must drastically reduce the scope and the cost of virtualization to make do with IoT hardware constraints.

The Femto-Container architecture therefore relies on ultra-light\-weight virtualization, as well as on a set of assumptions and features regarding an underlying RTOS, defined below.

\if{0}
% Could use a citation
The proposed architecture is based on the functions-as-a-service (FaaS) architecture used in cloud deployments.
Multiple provider-controlled environments are hosted and presented to the end user as opaque execution environment.
A user can deploy functions on the environment and which are executed based on external inputs to these functions.

This architecture is favored because it provides a number of beneficial properties in terms of code mobility and cyber-security:

\begin{enumerate}
\item \texttt{Hardware abstraction}: The functions are written in a portable programming language and can be deployed directly as code on the provided cloud platform.
\item \texttt{OS abstraction}: The deployed function is independent of the operating system % elaborate a bit here
\item \texttt{Isolation}: the FaaS environment provides a natural defence line with the isolated environment.
\end{enumerate}
\fi

\if{0}
The standard container runtime architecture (typically used in the cloud) is depicted in \autoref{fig:arch-design}.2. A provider-controlled operating system hosts one or more virtual machines sharing the same hardware resources. Each virtual machine (VM) virtualizes an OS maintained by a third-party. Each VM can host one or more (typically many) containers. This architecture is favored because it naturally provides the below properties, which are crucial in terms of code mobility and cyber-security:
\begin{enumerate}
\item \texttt{Hardware abstraction}: the VM abstracts the hardware on top of which software is running, facilitating code portability across different hardware architectures and configurations.
\item \texttt{OS abstraction}: the container further enhances code mobility by offering standardized access to OS services, which facilitates code portability across different OS. 
\item \texttt{Isolation}: the VM provides a natural defence line. If an attackers breaks out of a container, it only compromises the one VM it is in, other VMs and the host system are safe.
\end{enumerate}

However, the standard container runtime architecture depicted \autoref{fig:arch-design}.2 faces issues on low-power IoT devices with constrained resources:
\begin{itemize}
\item Full OS virtualization leads to a prohibitive toll on resources and/or execution speed;
\item Standard containers solutions are not applicable on constrained IoT devices.
\end{itemize}
\fi

%\subsection{Femto-Container Architecture}
%\label{sec:femto-arch-design}
%{\bf \large{Femto-Container Architecture.}}

\if{0}
For these reasons, we introduce Femto-Containers, an new runtime architecture targeting constrained IoT devices, as described in the following.
Instead of using traditional containers inside large VMs which virtualize a full-blown OS,
we host smaller VMs inside simplified containers, running on top of a real-time operating system (RTOS), as depicted in \autoref{fig:arch-design}.1.
By flipping around virtualization and containerization,
we can retain the crucial properties w.r.t. code mobility and cyber-security
(as we still combine isolation, hardware/OS abstraction)
but we are able to drastically reduce the scope of virtualization and its cost on constrained IoT devices.
\fi

\if{0}
We retain the crucial properties of FaaS concerning code mobility and cybersecurity (as we still combine isolation, hardware/OS abstraction)

In this section, we introduce Femto-containers, a new embedded runtime architecture targeting constrained IoT devices, as described in the following.

Similarly to a FaaS runtime, we deploy and execute small modules of virtualized logic as shown in \autoref{fig:arch-design}.1, hosted on top of an abstraction layer which offers both tight integration, standard facilities, as well as isolation for these modules, with respect to the underlying OS. 
By combining isolation and hardware/OS abstraction, we retain the typical properties of FaaS runtimes concerning code mobility and cybersecurity.

Differently from typical FaaS runtimes, however, Femto-containers must be able to interact with specific hardware (\emph{e.g.}, sensor/actuators), and must drastically reduce the scope and the cost of virtualization to make do with constrained IoT hardware constraints.

By offering a tight and standardised integration between the virtualization environment and the RTOS, we retain the crucial properties of FaaS concerning code mobility and cybersecurity (as we still combine isolation, hardware/OS abstraction) but we are able to drastically reduce the scope of virtualization and its cost on constrained IoT devices.

A distinct requirement for Femto-Container, as opposed to the cloud FaaS architecture, is the need to select the physical devices the functions are executed on.
Where FaaS assumes a homogeneous execution environment, in a wireless sensor network deployment this assumption can not be made.
Nodes within the network can have different sensors attached, depending on their location and perceived functionality.
Femto-Containers can be deployed on different nodes depending on the hardware available on the physical nodes.

Similar to the large scale FaaS architecture, the Femto-Container architecture thus aims to trade off features for a more lightweight approach suitable for small embedded devices and provide a generic code execution environment.
In more details, the Femto-Container architecture relies on a number of RTOS-provided features and a set of assumptions, as described below.
% Expand on the management size
\fi

\iffalse
\begin{figure}
\centering
\begin{subfigure}{.5\columnwidth}
  \centering
  \includegraphics[width=.8\linewidth]{Figures/femto-arch.png}
  \caption{Femto-container runtime : \\ Small VMs in containers.}
  \label{fig:sub1}
\end{subfigure}%
\begin{subfigure}{.5\columnwidth}
  \centering
  \includegraphics[width=.8\linewidth]{Figures/cloud-native-arch.png}
  \caption{Cloud container runtime : \\ Containers in large VMs.}
  \label{fig:sub2}
\end{subfigure}
\caption{Runtime architecture: Femto \textit{vs} Cloud.}
\label{fig:arch-design}
\end{figure}
\fi

\paragraph{\bf Use of an RTOS with Multi-Threading}
It is assumed that the RTOS supports real-time multi-threading with a scheduler.
Each Femto-Container runs in a separate thread.
Well-known operating systems in this space can provide for that, such as RIOT~\cite{baccelli2018riot} %\ysh{(TODO: why does this cite use et al? )}
or FreeRTOS~\cite{Barry:2012} and others~\cite{hahm2015operating}.
These  can run on the bulk of commodity microcontroller hardware available. 
Note that RTOS facilities for scheduling enable simple controlling of how Femto-Containers interfere with other tasks in the embedded system.

\paragraph{\bf No Assumptions on Microcontroller Hardware}
To retain generality, we aim for a purely software-based isolation, which can also run on the least capable microcontrollers, without any assumptions on hardware architecture enhancements or security peripherals.
If present, hardware-based isolation features could nevertheless be used to add layers of protection in-depth.
For instance TrustZone software module isolation relies on enhanced Arm Cortex-M microcontroller architectures~\cite{pinto2019trustzone}.
Other examples using hardware based protection mechanisms are TockOS~\cite{tockos} or Amulet~\cite{hardin2018mem-isolation-MCUs}, which rely on a hardware Memory Protection Unit (MPU) to isolate software modules.

\paragraph{\bf Use of Ultra-Lightweight Virtualization}
The virtual machine provides hardware agnosticism, and should therefore not rely on any specific hardware features or peripherals.
This allows for running identical application code on heterogeneous hardware platforms.
The virtual machine instances must have a low memory footprint, both in Flash and in RAM.
This allows to run multiple VMs in parallel on the device.
%Last but not least, VM execution slow-down must be tolerable compared to native execution.
Note that, since we aim to virtualize less functionalities, the VM can in fact implement a reduced feature set.
For instance, virtualized peripherals such as an interrupt controller are not required, and we give up the possibility of virtualizing a full OS. 
As exact virtualized hardware and peripherals is not a requirement, solutions around the interpretation of a scripting language are also suitable as target for Femto-Containers.

\subsubsection*{\bf Use of OS Interfaces}
A slim environment around the \ac{VM} exposes RTOS facilities to the \ac{VM}.
The container sandboxing a VM allows this VM to be independent of the underlying operating system, 
and provides the facilities as a generic interface to the \ac{VM}.
Simple contracts between container and RTOS can be used to define and limit the privileges of a container regarding its access to OS facilities.
%Code on the virtual machine must not be able to access resources outside the limitations set by the contracts and the OS.
Note that such limitations must be enforced at run-time to safely allow 3rd party module reprogramming.

\paragraph{\bf Isolation \& Sandboxing through Virtualization}
The OS and Femto-Containers must be mutually protected from malicious code, as described in \autoref{sec:isolationmodel}.
This implies in particular that code running in the \ac{VM} must not be able to access memory regions outside of what is granted via permissions.
Here again, simple contracts can be used to define and limit memory and peripheral access of the code running in the Femto-Container.
The strong isolation and security of the sandbox must prevent tenant escalation to the operating system and other tenants.

\paragraph{\bf Slim Event-based Launchpad Execution Model}
Femto-Containers are executed on-demand, when an event in the RTOS context calls for it.
Femto-Container applications are rather short-lived and have a finite execution constraint.
This execution model fits well with the characteristics of most low-power IoT software.
To simplify containerization and enforce security-by-design, we mandate that 
Femto-Containers can only be attached and launch from predetermined launch pads, which are sprinkled throughout the RTOS firmware.
Where applicable however, the result from the Femto-Container execution can modify the control flow in the firmware as defined in the launch pad.
%Note that these characteristics can both provide security-by-design and simplify containerization.
%However, an assumption is that adding (many) launch-pads is easy and incurs negligible resource consumption overhead.\\
%Each invocation launches a fresh Femto-Container instance, without retaining execution state between the invocations.\\
% Mention something on that the state of the vm is not persistent
% hooks sprinkled throughout the firmware, needs a full upgrade when modified
% Result of the container could affect the control flow after the hook
% preprovisioning is done in the firmware, adds low cost flexibility to the running firmware

% We do not want to run a full OS inside
%\subsubsection*{}
% Compared to the large scale cloud architecture, the Femto-Container trades features for a more lightweight approach suitable on small embedded devices. It is not designed to run on stand-alone bare metal devices but instead relies on a number of OS facilities provided. Furthermore, The reduced feature set in the virtual machine itself removes the option to run a full (embedded) OS inside a Femto-Container, but limits it to simple applications instead.

\paragraph{\bf Low-power Secure Runtime Update Primitives}
Launching a new Femto-Container or modifying an existing Femto-Container can be done without modifying the RTOS firmware. However, updating the hooks themselves requires a firmware update.
In our implementation, both types of updates use CoAP network transfer and software update metadata defined by SUIT~\cite{moran2019firmware} (CBOR, COSE) to secure updates end-to-end over network paths including low-power wireless segments~\cite{zandberg2019secure}.
Leveraging SUIT for these update payloads provides authentication, integrity checks and roll-back options.
Updating a Femto-Container application attached to a hook is done via a SUIT manifest.
The exact hook to attach the new Femto-Container to is done by specifying the hook as a  unique identifier (UUID) as a  storage location in the SUIT manifest.
A rapid develop-and-deploy cycle only requires a new SUIT manifest with the storage location specified every update.
Sending this manifest to the device triggers the update of the hook after the new Femto-Container application is downloaded to the device and stored in the RAM.
Using a secure update mechanism such as SUIT prevents a malicious client from intervening in the update and installation process of a Femto-Container application.

\section{Ultra-Lightweight VM Micro-Benchmarks}

In this section, we compare the performance of a proof of concept using RIOT~\cite{baccelli2018riot} to extend with Femto-Container functionality, based on different ultra-lightweight isolation techniques: Python (MicroPython runtime), WebAssembly (WASM3 runtime), eBPF (rBPF runtime) and Javascript (RIOTjs runtime). 

We run experiments using each virtualization candidate on an off-the-shelf IoT hardware platform representative of the landscape of modern 32-bit microcontroller architectures available: Arm Cortex-M4. Details of the benchmark setup are in Appendix A.

In the benchmarks we report on below, each implementation is loaded with a VM hosting logic performing a Fletcher32 checksum on a \SI{360}{\byte} input string. We reason that this computing load roughly mimics the instruction complexity of intensive sensor data (pre-)processing on-board.

\begin{table}[ht]
  \centering
  \begin{tabular}{lrr}
    \toprule
  Runtime  & ROM size & RAM size \\
    \midrule
    WASM3  & \SI{64}{\kibi\byte} & \SI{85}{\kibi\byte} \\
    rBPF   & \SI{4,4}{\kibi\byte} & \SI{0,6}{\kibi\byte} \\
%    rBPF Interpreter   & \SI{4440}{\byte} & \SI{660}{\byte} \\
    RIOTjs        & \SI{121}{\kibi\byte} & \SI{18}{\kibi\byte} \\
    MicroPython        & \SI{101}{\kibi\byte} & \SI{8,2}{\kibi\byte} \\
%    Femto-Containers   & \SI{4641}{\byte} & \SI{664}{\byte} \\ % TODO: Move to later benchmarks
    \midrule
    Host OS (without VM)   & \SI{52,5}{\kibi\byte} & \SI{16,3}{\kibi\byte} \\
%    Host OS (without VM)   & \SI{52760}{\byte} & \SI{14856}{\byte} \\

    \bottomrule
  \end{tabular}
  \vspace{0.5em}
\caption{Memory requirements for Femto-Container runtimes.\\}
\label{tbl:vmsize}
  \vspace{-1.5em}
\end{table}

\begin{table}[ht]
  \centering
  \begin{tabular}{lrrr}
    \toprule
   Runtime & code size & cold start overhead&run time \\
    \midrule
    Native C         & \SI{74}{\byte}  & -- &\SI{27}{\micro\second} \\
    WASM3            & \SI{322}{\byte} & \SI{17096}{\micro\second} &\SI{980}{\micro\second} \\
    rBPF             & \SI{456}{\byte} & \SI{1}{\micro\second} & \SI{2133}{\micro\second} \\
%    Femto-Containers & \SI{456}{\byte} & \SI{}{\micro\second} & \SI{2148}{\micro\second} \\  % TODO: Move to later benchmarks
    RIOTjs      & \SI{593}{\byte} & \SI{5589}{\micro\second} & \SI{14726}{\micro\second} \\
    MicroPython      & \SI{497}{\byte} & \SI{21907}{\micro\second} & \SI{16325}{\micro\second} \\
    \bottomrule
  \end{tabular}
  \vspace{0.5em}
  \caption{Size and performance of fletcher32 logic hosted in different Femto-Container runtimes on Cortex-M4.}
  \label{tbl:scriptsize}
  \vspace{-1.5em}
\end{table}

Our benchmarks results are shown in Table~\ref{tbl:vmsize} and Table~\ref{tbl:scriptsize}.
The startup time measures the time it takes for the runtime to load the application.
This contains setup processing to parse the application or JIT compilation steps.
%The results highlight how much the footprint of hosting logic in a VM can vary, depending on the virtualization technique being used.

\paragraph*{\bf Looking at size}
While the size of applications are roughly comparable accross virtualization techniques (see Table~\ref{tbl:scriptsize}), the memory required on the IoT device differs wildly.
In particular, techniques based on script interpreters (RIOTjs and MicroPython) require the biggest dedicated ROM memory budget, above \SI{100}{\kibi\byte}.

For comparison, the biggest ROM budget we measured requires 27 times more memory than the smallest budget.
Similarly, RAM requirements vary a lot. Note that we could not determine with absolute precision the lower bound for script interpreters techniques, due to some flexibility given at compile time to set heap size in RAM. Nevertheless, our experiments show that the biggest RAM budget requires 140 times more RAM than the smallest budget.
We remark that, as noted in prior work~\cite{zandberg2020minimal} the minimum required page size of \SI{64}{\kibi\byte}  to comply with the WebAssembly specification explains why WASM3 performs poorly in terms of RAM. One can envision enhancements where this requirement is relaxed. However the RAM budget would still be well above an order of magnitude more than the RAM budget we measured with rBPF.

Last but not least, let's give some perspective by comparison with a typical memory budget for the \emph{whole} software embedded on the IoT device.  As a reminder, in the class of devices we consider, a microcontroller memory capacity of 64kB in RAM and 256kB in Flash (ROM) is not uncommon. A typical OS footprint for this type of device is shown in the last row of Table~\ref{tbl:vmsize}. For such targets, according to our measurements, adding a VM can either incur a tremendous increase in total memory requirements (200\% more ROM with MicroPython) or a negligible impact (8\% more ROM with rBPF) as visualized in \autoref{fig:candidates-memory-distrib}.

\paragraph*{\bf Looking at speed}
To no surprise, beyond size overhead, virtualization also has a cost in terms of execution speed.
But here again, performance varies wildly depending on the virtualization technique.
On one hand, solutions such as MicroPython and RIOTjs directly interpret the code snippet and execute it.
On the other hand, solutions such as rBPF and WASM3 require a compilation step in between to convert from human readable code to machine readable.

Our measurements show that script interpreters incur an enormous penalty in execution speed.
Compared to native code execution, script interpreters are a whooping ~600 times slower.
Compared to the same base (native execution) WASM is only 37 times slower, and rBPF 77 times slower.

One last aspect to consider is the startup time dedicated to preliminary pre-processing when loading new VM logic, before it can be executed (including steps such as code parsing and intermediate translation, various pre-flight checks etc.).
Depending on the virtualization technique, this startup time varies almost 1000 fold -- from a few microseconds compared to a few milliseconds.

\begin{figure*}[t]
%  \centering

% Measured examples/suit-update on the nrf52840dk with libcose and C25519 for crypto. Disabled is the shell-commands, ethos and the progress bar
% Crypto: 4708 + 895 + 1302 + 256  = 7161
% Network: 19226 + 234 + 60 + 8 + 585 = 20113
% Kernel: 6474 + 2640 + 2128 + 124 + 52 + 16 + 172 + 752 + 572 +  350 +334 + 212 + 140 + 130 + 74 + 58 + 56 + 38 + 34 + 28  + 2655 = 17039
% OTA: 1258 + 5006 + 546 + 250 + 568 + 232 + 312 = 8172
% Total size: 7161 + 20113 + 17039 + 8172 = 52485
\iffalse
\begin{subfigure}{\columnwidth}
    \begin{tikzpicture}[scale=0.7]
 \pie[explode=0.1, text=inside, radius=2.5]
    {14/,
     38/,
     32/,
     16/
     }
    \end{tikzpicture}
    \caption{RIOT without hosting engine(53kBytes).}
  \end{subfigure}
\fi
% Total size: 7161 + 20113 + 17039 + 8172 + 101000 = 153485
   \begin{subfigure}{\columnwidth}
    \begin{tikzpicture}[scale=0.7]
 \pie[explode=0.1, text=inside, radius=2.5]
    {5/,
     13/,
     11/,
     5/,
     66/
     }
    \end{tikzpicture}
    \caption{RIOT with MicroPython Femto-Container (154kBytes).}
  \end{subfigure}
% Total size: 7161 + 20113 + 17039 + 8172 + 4440 = 56925
  \begin{subfigure}{\columnwidth}
    \begin{tikzpicture}[scale=0.7]
 \pie[explode=0.1, text=legend, radius=2.5]
    {13/Crypto,
     35/Network stack,
     30/Kernel,
     14/OTA module,
     8/Femto-Container runtime
     }
    \end{tikzpicture}
    \caption{RIOT with rBPF Femto-Container (57kBytes).}
  \end{subfigure}
  \caption{Flash memory distribution with different Femto-Containers. RIOT is configured with 6LoWPAN, CoAP, SUIT-compliant OTA (totalling 53kBytes in Flash memory).}
  \label{fig:candidates-memory-distrib}
\end{figure*}

%Dismiss micropython based on impact on the binary size. Allows for rapid prototyping, but impact on flash too large to be relevant here. Maybe also too slow to consider for debugging scripts.

%For the WASM3 WebAssembly interpreter, both the flash footprint and the required RAM is huge, the last one due to the page \SI{64}{\kibi\byte} requirement of the WebAssembly specification.

%In terms of security, discuss architecture? Discuss LoC ? (Perspecive would be to formally verify the hosting engine).

%\subsection{Perspectives}

%We now aim to flesh out our Femto-Container implementation based on a particular virtualization technique. In this section, we reason our choice of \acp{VM} to efficiently address our target use cases (described in Section~\ref{sec:case-study}), which involve hosting and mutually isolating multiple VMs which may contain either high-level business logic, or low-level debug/monitoring code snippets.

\paragraph*{\bf Considering VM architecture \& security}
%There are notable architectural differences amongst the solutions we pre-selected and looked at in this this section. For instance, 
WASM, MicroPython and RIOTjs each require some form of heap on which to allocate application variables.
On the other hand, rBPF does not require a heap.
With a view to accommodating several VMs concurrently, a heap-based architecture presents on the one hand some potential advantages in terms of memory (pooling) efficiency, but on the other hand some potential drawbacks in terms of security with mututal isolation of the VMs' memory between different tenants.

Furthermore, security guarantees call for a formally verified implementation of the hosting engine down the road.
A typical approximation is: less lines of code (LoC) means much less effort to produce a verified implementation.
For instance, the rBPF implementation is ~1,5k LoC, while the WASM3 implementation is ~10k LoC.
The other implementations we considered in our pre-selection, RIOTjs and MicroPython, encompass significantly more LoC.

\subsection{Choice of Virtualization}
Our benchmarks indicate that in terms of memory overhead, startup time and LoC, Femto-Containers using eBPF virtualization is the most attractive, by far.
We note that execution time with WebAssembly is 2x faster than with rBPF. However, we expect that a 2x factor in execution time will have no significant impact in practice for the use cases we target, such as small processing workloads.
Since our priority is on memory footprint (recall our aim of $\approx$10\% memory overhead for functionality containerization) we thus choose rBPF to flesh out our concept further. 
%These results both extend and confirm independent results presented in prior work~\cite{zandberg2020minimal}.or the use cases we target, 

%\pagebreak
\section{Femto-Container Runtime Implementation}
%The Femto-Containers virtual machine design is based on a
% is designed from the eBPF instruction set architecture.
%The instruction set itself is minimal and optimized for fast parsing with compact code.
%The virtual machine hosting engine interprets the eBPF instructions and updates the virtual machine state.
%The Femto-Container architecture described above is implemented.
As proof of concept, we implemented the Femto-Container architecture, with functions hosted in the operating system RIOT and virtualization using an instruction set compatible with the eBPF instruction set.
This implementation is open source (published in \cite{femto-github}). We detail below its main characteristics.

\if{0}
% Move this to the relevant architecture, keep this high level
The Femto-Container engine is designed to host and interpret one or more small virtual machines, hosted on top of an OS, running on a microcontroller-based IoT device.
The virtual machine instruction set is compatible with the eBPF instruction set architecture.
The address space inside the virtual machine is shared with the operating system, no memory address translation is used by default.
% Mention address space sharing
\fi

\paragraph*{Simple Containerization}
Simplified interfaces provide a uniform environment around the VM, independent of the RIOT operating system.
%However this requires a number of facilities to be provided by the RTOS to the RTOS-specific Femto-Container implementation.
Access from the Femto-Container to the required OS facilities is allowed through system calls to services provided by RIOT.
These system calls can be used by the loaded applications via the eBPF native \texttt{call} instruction.
Furthermore, the OS can share specific memory regions with the container.
\ \\
{\bf Key-value store.}
%The Femto-Containers ecosystem provided with the host OS includes a simple key value store.
In lieu of a file system, applications hosted in Femto-Containers can load and store simple values, by a numerical key reference, in a key-value store.
This provides a mechanism for persistent storage, between application invocations.
Interaction with this key-value store is implemented via a set of system calls, keeping it independent of the instruction set.
By default, two key-value stores are provided by the OS.
The first key-value store is local to the application, for values that are private to the VM accommodated in the container.
The second key-value store is global, and can be accessed by all applications, used to communicate values between applications.
An optional third intermediate-level of key-value store is possible to facilitate sharing data across a set of VMs from the same tenant, while isolating this set of VMs from other tenants' VMs.

% OS integration
%\paragraph{RTOS Integration}
\paragraph*{\bf Use of RIOT Multi-Threading}
%The RTOS is used to run Femto-Containers and provide a number of services to the hosting engine.
%Within the RTOS, the Femto-Container instances are scheduled as regular threads.
Each Femto-Container application instance running is scheduled as a regular thread in RIOT.
The native OS thread scheduling mechanism can simply execute concurrently  and share resources amongst multiple Femto-Containers and other tasks, spread over different threads.
A Femto-Containers instance requires minimal RAM: a small stack and the register set, but no heap.
The host RTOS bears thus a very small overhead per Femto-Containers instance.

\begin{figure}[t]
  \centering
    \includegraphics[width=.8\columnwidth]{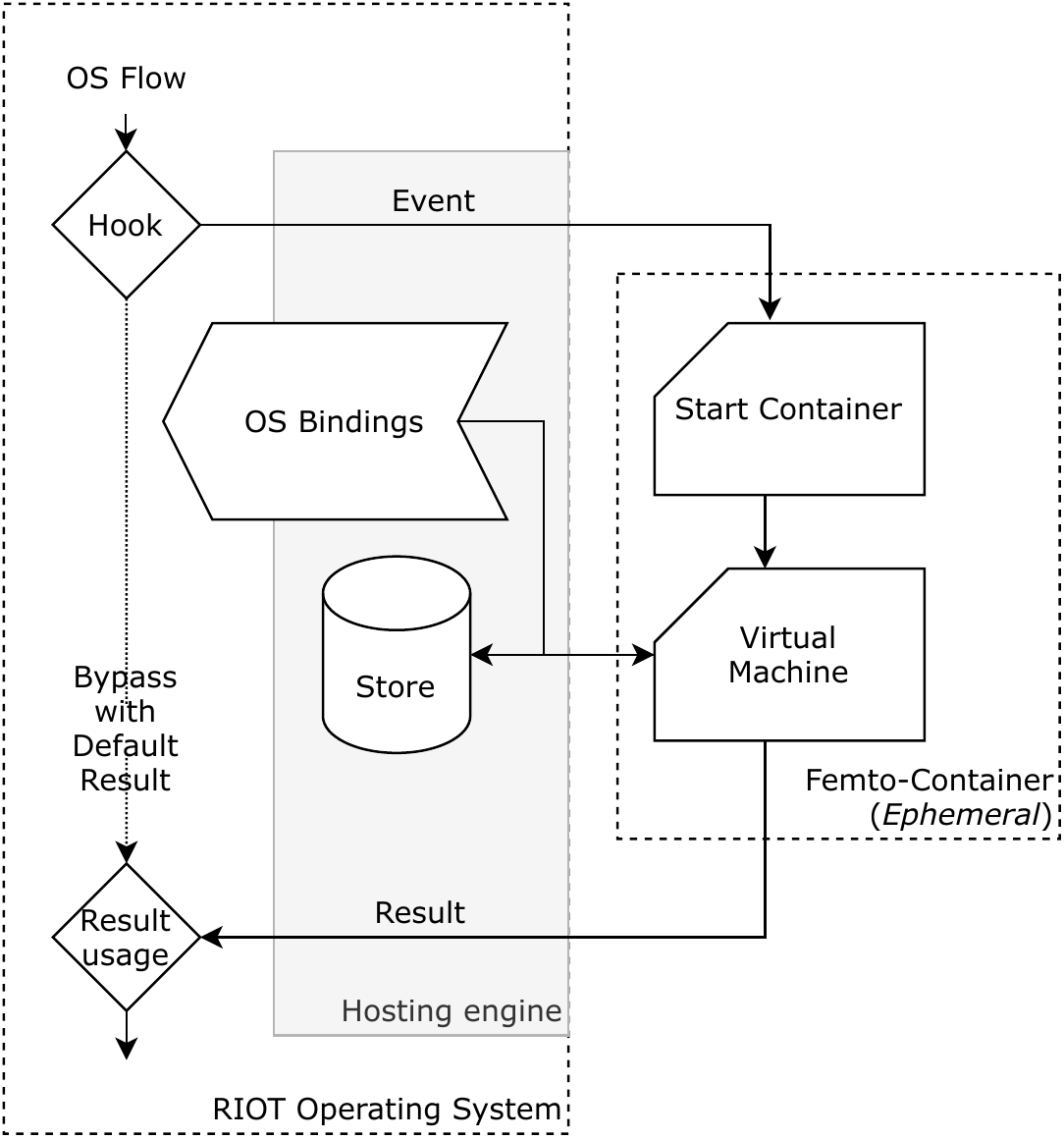}
  \caption{Femto-Container RTOS integration.}
  \label{fig:vm-overview}
\end{figure}

\autoref{fig:vm-overview} shows how Femto-Containers integrate into the operating system.An overview of how Femto-Containers integrate in the operating system is shown in \autoref{fig:vm-overview}.
It shows the flow within the operating system, with the Femto-Container triggered by an event in the operating system.
The Femto-Container is started if it is present and gains access to it's own store and any bindings allowed by the operating system.

As the Femto-Container Instance does not virtualize its own set of peripherals, no interrupts or pseudo-hardware is available to the Femto-Container application. This also removes the option to interrupt the application flow inside a Femto-Container.

The  hardware and peripherals available on the device are not accessible by the Femto-Containers instances.
All interaction with hardware peripherals passes through the host RTOS via the system call interface.

\paragraph*{\bf Ultra-Lightweight Virtualisation using eBPF}% Instruction Set}
Application code is virtualized using Femto-Containers, our enhancement of the rBPF virtual machine implementation.
rBPF is again based on the Linux eBPF.
The architectures of these virtual machines are similar enough that they all use the LLVM compiler with the eBPF target for compilation.
%instruction set to microcontrollers, which we achieved in a way akin to rBPF~\cite{zandberg2020minimal}). \todoEB{Any difference we could mention with rBPF?}
%
%From the CAV paper
%\jp{{rBPF} uses a reduced, 64-bit sized, instruction set architecture (ISA), fixed-sized 512 bytes stack and no heap interaction. It has no special registers or interrupts for flow control, or virtual memory: the host device's memory is only guarded by permissions.}
%It has the minimal architecture without peripherals required and is designed with \ac{VM} applications in mind.
\\
% Remove (or move to section 4.3)
{\bf Register-based VM.} The virtual machine operates on eleven registers of 64 bits wide.
The last register (\texttt{r10}) is a read-only pointer to the beginning of a \SI{512}{\byte} stack provided by the femto-container hosting engine.
Interaction with the stack happens via load and store instructions.
%as shown in \autoref{fig:isa}.
% moving values from the registers to the stack and vise versa.
Instructions are divided into an \SI{8}{\bit} opcode, two \SI{4}{\bit} registers: source and destination, an \SI{16}{\bit} offset field and an \SI{32}{\bit} immediate value.
%Unused fields in an instruction are set to zero.
Position-independent code is achieved by using the reference in \texttt{r10} and the offset field in the instructions.

\iffalse
\begin{figure}[t]
  \centering
    \includegraphics[width=\columnwidth]{Figures/isa.pdf}
  \caption{eBPF instructions, registers and stack.}
  \label{fig:isa}
\end{figure}
\fi
{\bf Jumptable \& Interpreter.}
The interpreter parses instructions and executes them operating on the registers and stack.
The machine itself is implemented as a computed jumptable, with the instruction opcodes as keys.
During execution, the hosting engine iterates over the instruction opcodes in the application,
and jumps directly to the instruction-specific code.
This design keeps the interpreter itself small and fast.

\iffalse
% Koen: removed this paragraph, doesn't fit the story anymore
{\bf Ahead-of-Time vs Just-in-Time.}
One approach to speed up embedded execution time is to perform a translation into device-native code.
One way to offload the device is to use more Ahead-of-Time (AoT) compilation/interpretation, and less Just-in-Time (JiT) processing on-device. 
However, using AoT pre-compiled code can both complicate run-time security checks on-board the IoT device, and reduce the portability of the code deployed on the device.
For these reasons, in this paper, we consider primarily JiT.
\fi

\subsection*{Isolation \& Sandboxing}
To control the capabilities of Femto-Containers, and to protect the OS from memory access by malicious applications, a simple but effective memory protection system is used.
By default each virtual machine instance only has access to its \ac{VM}-specific registers and its stack.

\begin{figure}[t]
  \centering
    \includegraphics[width=\columnwidth]{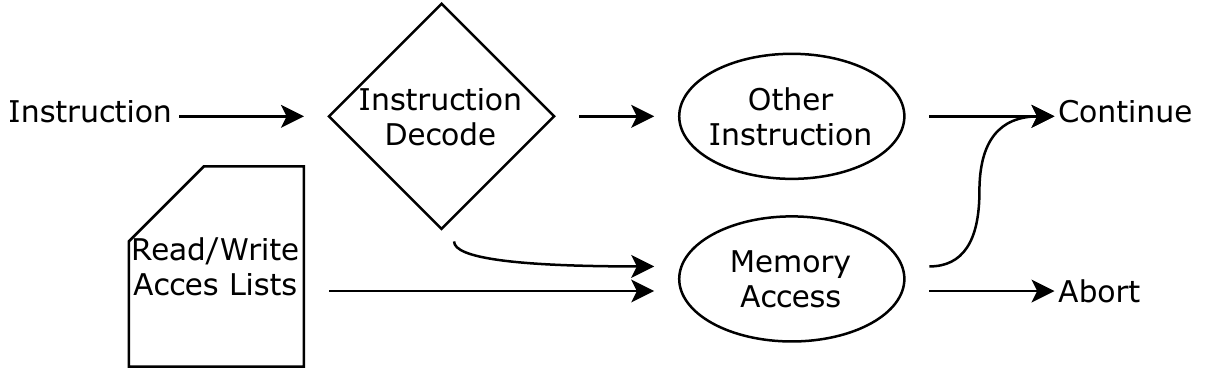}
  \caption{Interaction between memory instructions and the access lists.}
  \label{fig:memaccess}
\end{figure}

{\bf Memory access checks at runtime.}
Allow lists can be configured (attached in the hosting engine) to explicitly allow a \ac{VM} instance access to other memory regions.
%In the hosting engine this is implemented by attaching a set of memory regions to the \ac{VM} instance.
These memory regions can have individual flags for allowing read/write access.
%With this the virtual machine can be granted access to data outside the virtual machine.
For example, a firewall-type trigger can grant read-only access to the network packet, allowing the virtual machine to inspect the packet, but not to modify it.
As the memory instructions allow for calculated addresses based on register values, memory accesses are checked at runtime with the access lists against the resulting address, as show in \autoref{fig:memaccess}. Illegal access aborts execution.
While this relies on properly configured allow lists, it prevents access to memory outside the sandbox by a malicious tenant.
%As shown in \autoref{fig:memaccess}, when the memory instruction is executed, the list of allowed memory regions is consulted and either the access is allowed (and execution continues), or the \ac{VM} execution is aborted because of an illegal memory access.

{\bf Pre-flight instruction checks.} \label{sec:pre-flight}
A Femto-Container verifies the application before it is executed for the first time.
These checks include checks on the individual instruction fields.
For example, as there are only 11 registers, but space in the instruction for 16 registers, the register fields must be checked for out-of-bounds values.
A special case here is register \texttt{r10} which is read-only, and thus is not allowed in the destination field of the instructions.

The jump instructions are also checked to ensure that the destination of the jump is within the address space of the application code.
As calculated jump destinations are not supported in the instruction set, the jump targets are known before executions and are checked during the pre-flight checks.
During the execution of the application, the jump destinations no longer have to be verified and can be accepted as valid destinations.
This prevents a tenant from jumping execution to the application code outside of the sandbox such as code from a different tenant or vulnerable code planted otherwise.

Finite execution is also enforced, by limiting both the total number of instructions $ N_i $, and the number of branch instructions $ N_b $ that are allowed.
In practice, this limits the total number of instructions executed to:
$ N_i \times N_b $.
This puts a limit on the computational resource exhausted by a single execution by putting a hard limit on the total number of instructions executed.

\subsection*{Hooks \& Event-based Execution}
The Femto-Container hosting engine instantiates and runs containers as triggered by events within the RTOS.
%A short-lived application is executed based on an event in the operating system.
Such events can be a network packet reception, sensor reading input or an operating system scheduling events for instance.
Business logic applications can be implemented either by directly responding to sensor input or by attaching to a timer-based hook to fire periodically.

Simple hooks are pre-compiled into the RTOS firmware, providing a pre-determined set of pads from which Femto-Containers can be attached and launched.

\begin{lstlisting}[language=c,label=lst:hook,caption={Example hook implementation.}]
sched_ctx_t context = {
    .previous = active_thread,
    .next = next_thread,
};

int64_t result;

f12r_hook_execute(FC_HOOK_SCHED, &context,
                sizeof(context), &result);
\end{lstlisting}

An example of a hook integrated in the firmware is shown in \autoref{lst:hook}.
The firmware has to set up the context struct for the Femto-Containers after which it can call the hosting engine to execute the Femto-Container instances associated with the hook.

% Mention something on that the state of the vm is not persistent
% hooks sprinkled throughout the firmware, needs a full upgrade when modified
% Result of the container could affect the control flow after the hook
% preprovisioning is done in the firmware, adds low cost flexibility to the running firmware

\section{Use-Case Prototyping with Femto-Containers}
\label{sec:femto-container-examples}
In this section, we describe and demonstrate the programming model exposed by Femto-Containers.
We use Femto-Containers to prototype the implementation of several use cases involving one or more functions (applications).
Where multiple functions are involved, these are hosted concurrently on a single microcontroller.
The goal of these functions is to match the scenarios we targeted initially in \autoref{sec:case-study}.

%\jp{I assume this regards C program compiled to rBPF?}
In the prototype implementation we show below, we used C to code the applications hosted on Femto-Containers engine.
However, any other target language supported by LLVM could be used instead such as C++ and Rust, for instance.

\subsection{Programming Model}
\label{sec:femto-programming-model}
%
%\todoEB{Describe what application / programming model is exposed by the container.}

Femto-Containers follow an event-driven programming model.
Applications hosted are only executed when triggered by events in the operating system.
The applications specify the entry point and to which hooks they are to be attached inside the operating system.

The logic that can be deployed in Femto-Containers is limited following the eBPF architecture.
For instance, asynchronous operation is not supported: there is no option to interrupt the control flow inside a Femto-Container application from outside the virtual machine.
This is mainly caused by the simple nature of the architecture: neither interrupts nor indirect jumps are available.
This trade-off reduces flexibility but increases the security for the host operating system.

Our current implementation is also limited by the fixed, small size of the stack (512 Bytes) dictated by the eBPF specification.
% which limits the available memory to Femto-Container applications.
More memory-consuming tasks would need special handling to provide additional memory.
An enhanced implementation could however allow the application to request more stack from the RTOS, for example via the contracts, which would alleviate this issue.
More computation- and memory-intensive tasks could also make use of additional system calls provided by the RTOS, which could execute generic primitives at native speed.

The Femto-Container hosting engine is designed to be fast enough to start applications on a hot code path without affecting significantly normal operating system execution times.
Small applications can thus be inserted and execute without substantial overhead or slowdown for the RTOS.
This way small debug or inspection applications can be inserted into the RTOS at any point as long as the application does not cause a deadline to be exceeded in the RTOS.

%With these properties, Femto-Containers is optimized for short-lived and event-driven applications.
%These can be executed often and included in the hot path of operating system code.
% Elaborate a bit later on

\subsection{Kernel Debug Code Example}
\label{sec:femto-debug-example}
The first prototype we display consists in a single application, which intervenes on a hot code path:
it is invoked by the scheduler of the OS.
It keeps an updated count of each threads' activations.
The logic hosted in the Femto-Container is shown in \autoref{lst:thread_log}.
A small C struct is passed as context, which contains the previous running thread ID and the next running thread ID.
The application maintains a value for every thread, incremented every time the thread is scheduled.
External code can request these counters and provide debug feedback to the developer.
%When compiled to bytecode and loaded into the femto-container hosting engine, the application is \SI{104}{\byte} in size.

\begin{lstlisting}[language=c,label=lst:thread_log,caption={Thread counter code.}]
#include <stdint.h>
#include "bpf/bpfapi/helpers.h"

#define THREAD_START_KEY    0x0

typedef struct {
    uint64_t previous; /* previous thread */
    uint64_t next;     /* next thread */
} sched_ctx_t;

int pid_log(sched_ctx_t *ctx)
{
 /* Zero pid means no next thread */
    if (ctx->next != 0) {
        uint32_t counter;
	uint32_t thread_key = THREAD_START_KEY +
                ctx->next;
        bpf_fetch_global(thread_key,
                         &counter);
        counter++;
        bpf_store_global(thread_key,
                         counter);
    }
    return  0;
}
\end{lstlisting}

\subsection{Networked Sensor Code Example}
\label{sec:femto-sensor-example}
\begin{figure}[t]
  \centering
    \includegraphics[width=\columnwidth]{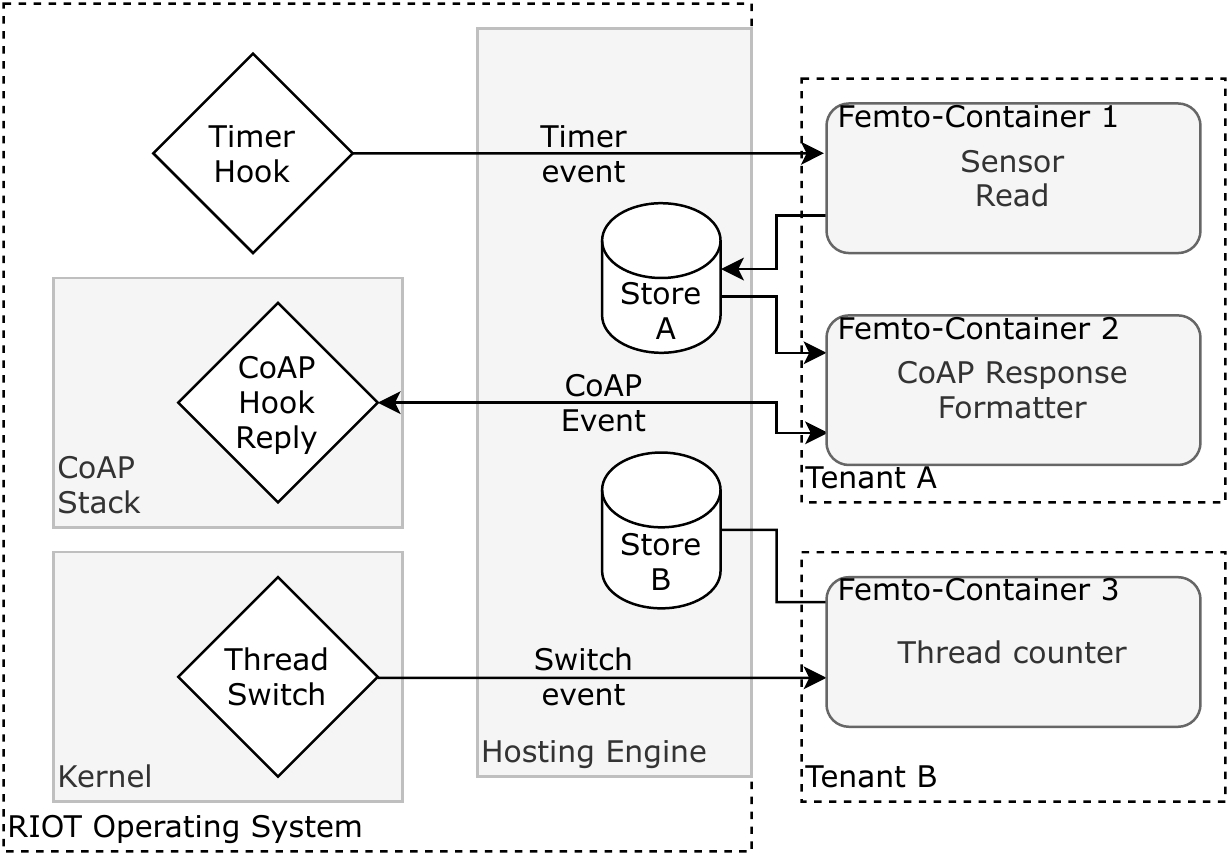}
  \caption{Event and value flow when hosting multiple containers from different tenants.}
  \label{fig:inter-vm}
\end{figure}

The second prototype we display adds two Femto-Containers from another tenant to the setup of the first prototype.
Interaction between these two additional containers is achieved via a separate key-value store, as depicted in \autoref{fig:inter-vm}.
The logic hosted in the first Femto-Container, periodically triggered by the timer event, reads, processes and stores a sensor value.
The code for this logic is shown in \cite{femto-sensor-process}.
The second container's logic is triggered upon receiving a network packet (CoAP request), and returns the stored sensor value back to the requestor.
The code for this logic is shown in \cite{femto-fetch}. %\ysh{TODO: should we give an anonymous repo instead of a cite?}

In this toy example, the sensor value processing is a simple moving average, but more complex post-processing is possible instead, such as differential privacy or some federated learning logic, for instance.
%such as machine learning.
This example sketches both how multiple tenants can be accommodated, and how separating the concerns between different containers is achieved (between sensor value reading/processing on the one hand, and on the other hand the communication between the device and a remote requester).

\section{Femto-Container Formal Verification}

\begin{figure*}[t!]
  \centerline{\includegraphics[width=\textwidth]{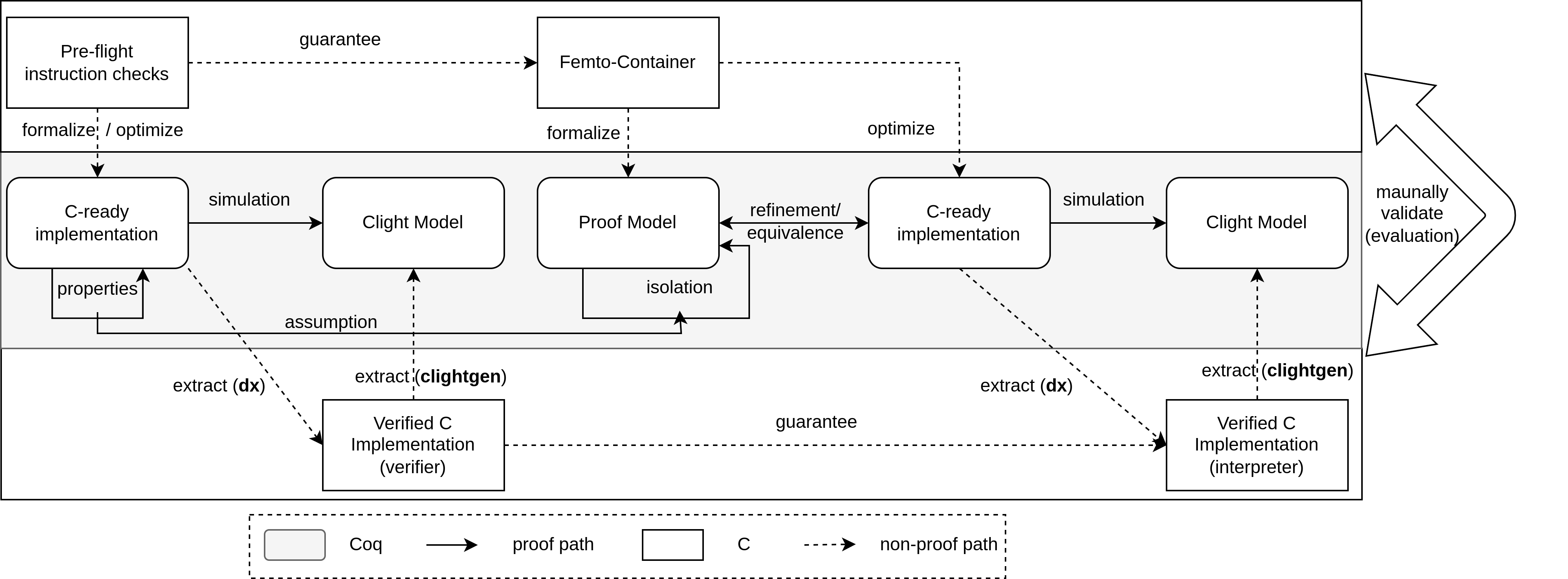}}
  \caption{{\cfc} Formal verification workflow.}
  \label{fig:verif-overview}
\end{figure*}

The critical components of Femto-Containers in terms of cyber-security are the rBPF interpreter and the pre-flight instruction checker. Since the implementation is conveniently small (500 lines of C code for the interpreter and the checker), we aimed at producing a formally verified implementation of these components. We will refer to \cfc\ (for Certified Femto-Container) as the runtime which uses the formally verified interpreter and checker.

\paragraph{Targeted requirements formalization.}
The security guarantees we wish to provide Femto-Containers with are essentially memory and fault isolation. More precisely, we want to prove it impossible for \cfc\ to access a memory location out of its app's register memory or to execute an instruction leading to an undefined behavior, and consequently heading the VM and/or its host to crash.
Providing these guarantees further strengthens the security needed with the threat model to prevent access to memory outside of the sandbox, in turn preventing unprivileged access to the operating system or other virtual machines.

\paragraph{Formal verification approach.}
We have used the Coq proof assistant to mechanically and exhaustively verify these requirements by employing the design workflow depicted in \autoref{fig:verif-overview}:
\begin{enumerate}[1)]
\item  First, we provide a proof model and a C-ready implementation that formalize (resp. optimize) the native, vanilla, C implementations of the rBPF verifier and virtual machine in RIOT.  Proof and "C-ready" models are proved semantically equivalent in Coq.

\item The verification of expected safety and isolation properties is performed by the Coq proof assistant on the VM's proof model. It relies on the formalized isolation guarantees provided by a) the CompCert C memory model \cite{Compcert}  ii) the pre-flight runtime checks of the verifier, and iii) the defensive runtime checks of the virtual machine itself (for numerical and memory operations).

\item The verified C implementation is automatically extracted from the C-ready Coq model using the $\partial x$ tool \cite{PIP}. Based on a set of formalized translation rule from Coq to C, $\partial x$ allows to craft a both reviewable and optimized C code from a functional Coq definition.

\item To ensure that the extracted C code refines the proof model, and hence satisfies the safety and isolation properties, the final simulation proof proceeds in two steps. %is performed in Coq by:
%
%\begin{enumerate}[a)] 
%\item 
First, a CompCert Clight model is extracted from the generated C code, using the VST-clightgen tool \cite{plcc}.
%
%\item and,
Second, proving that Clight model to simulate the C-ready model using translation validation \cite{Pnueli}.
\end{enumerate}

%\jp{after review this seems a fair summary, we could add more details from the commented text but they may be too technical for the Middleware croud :)}

\paragraph{Formal verification details.}
The Coq models and proofs used to obtain \cfc\ are presented in \cite{cavpaperanonymized} and available from \cite{cavartifactanonymized}.

\section{Performance Evaluation}

In this section we evaluate and compare the performance of Femto-Containers with rBPF and \cfc\ runtimes.
The comparison is done on a number of low-power IoT hardware platforms: Cortex-M4, RISC-V and ESP32 based microcontrollers.

\subsection{Hosting Engine Analysis}

We benchmark the Femto-Container implementation on a number of aspects.
First, we compare the footprint of the hosting engine on the embedded device.
This shows the impact of adding Femto-Containers to the applications
Second, we compare the execution time of a number of individual instructions.

%\todoEB{Analysis of code size etc. Stack size, speed etc.}
%\paragraph{Impact of pre-flight verification stages}:
%The number of pre-flight checks compared to rBPF has been increased. In rBPF the pre-flight checks are limited, the size is checked for a whole number of instructions and the last instruction must be a return instruction.
%In this section, we evaluate the footprint and the speed of execution with a Femto-Container, on various 32-bit microcontrollers.

\pgfplotstableread[col sep=comma,]{Measurements/femto_sizes.csv}\femtosizes
\pgfplotstableread[col sep=comma,]{Measurements/rbpf_sizes.csv}\rbpfsizes
\pgfplotstableread[col sep=comma,]{Measurements/certfc_sizes.csv}\certfcsizes
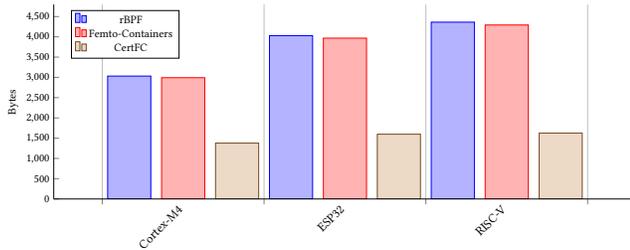
\begin{figure}[t!]
\captionsetup{aboveskip=-0.2\normalbaselineskip}

    \centering
    \resizebox{\columnwidth}{!}{
\begin{tikzpicture}
  {\centering
  \begin{axis}[
        ybar, axis on top,
        axis x line*=bottom,
        axis y line*=left,
        height=7cm, width=\textwidth,
        enlarge x limits=true,
        ylabel={Bytes},
        ytick distance=500,
        ymin=0,
        ybar interval=0.8,
        xtick=data,
        xticklabels = {{Cortex-M4}, {ESP32}, {RISC-V}},
        xticklabel style={font=\large, rotate=45, anchor=east},
        legend style={
            anchor=north west,
            legend pos=north west,
        },
    ]
    \addplot table [x expr=\coordindex, y index=1]{\rbpfsizes};
    \addplot table [x expr=\coordindex, y index=1]{\femtosizes};
    \addplot table [x expr=\coordindex, y index=1]{\certfcsizes};
    \legend{rBPF, Femto-Containers, CertFC};
  \end{axis}}
  \end{tikzpicture}
  }
  \caption{Flash requirement for the different implementations and platforms}
  \label{fig:femtosize}
\end{figure}

To compare the impact of adding the Femto-Containers to an existing firmware, we compare the memory footprint of the implementations.
In general, each Femto-Container needs memory to
\begin{itemize}
\item store the application bytecode;
\item handle the virtual machine state and stack.
\end{itemize}
The impact on the required flash on the firmware is shown in \autoref{fig:femtosize} and \autoref{tbl:femtosize}.
In terms of required RAM for execution, both rBPF and Femto-Containers show comparable flash and RAM memory usage.
In terms of Flash memory size, our measurements show that \cfc\ actually reduces the footprint by \SI{55}{\percent} on Cortex-M4.
The \cfc\ implementation requires slightly more memory, an increase of around \SI{50}{\byte} per instance.
This is caused by \cfc\ storing extra state of the virtual machine in the context struct and not on the thread stack.

\begin{table}[ht]
  \centering
  \begin{tabular}{lrr}
    \toprule
    & ROM size & RAM size \\
    \midrule
    Femto-Containers   & \SI{2992}{\byte} & \SI{624}{\byte} \\ % TODO: Move to later benchmarks
    rBPF   & \SI{3032}{\byte} & \SI{620}{\byte} \\
    \cfc & \SI{1378}{\byte} & \SI{672}{\byte} \\
    \bottomrule
  \end{tabular}
  \vspace{0.5em}
\caption{Memory footprint of a Femto-Container hosting minimal logic on Arm Cortex-M4.}
\label{tbl:femtosize}
\end{table}

The different implementations of Femto-Containers are compared in \autoref{fig:instructions2} against a set of eBPF instructions, showing that the rBPF extensions incur minimal overhead on the virtual machine and provide similar throughputs. %
%Visible in  is the instruction throughput of a number of instruction types available in the architecture.
%As is visible, there is almost no difference between rBPF and Femto-Containers.
Now,  the performance of the formally verified \cfc \; is lagging behind the other implementations, revealing the trade off between the formally verified code and a natively optimized implementation.

\pgfplotstableread[col sep=comma,]{Measurements/headers.dat}\headers
\pgfplotstableread[col sep=comma,]{Measurements/nrf52dk_rbpf_unit.csv}\nrfrbpfunit
\pgfplotstableread[col sep=comma,]{Measurements/nrf52dk_femto_unit.csv}\nrffemtounit
\pgfplotstableread[col sep=comma,]{Measurements/nrf52dk_certrbpf_unit.csv}\nrfcertrbpfunit

\begin{figure}[t!]
\captionsetup{aboveskip=-0.2\normalbaselineskip}

    \centering
    \resizebox{\columnwidth}{!}{
\begin{tikzpicture}
  {\centering
  \begin{axis}[
        ybar, axis on top,
        axis x line*=bottom,
        axis y line*=left,
        height=8cm, width=\textwidth,
        enlarge x limits=true,
        ylabel={\si{\micro\second} per instruction},
        ytick distance=0.25,
        ymin = 0,
        xtick=data,
        xticklabels from table={\headers}{test},
        xticklabel style={font=\large, rotate=45, anchor=east},
        legend style={
            anchor=north west,
            legend pos=north west,
        },
    ]
    \addplot table [x expr=\coordindex, y index=2]{\nrfrbpfunit};
    \addplot table [x expr=\coordindex, y index=2]{\nrffemtounit};
    \addplot table [x expr=\coordindex, y index=2]{\nrfcertrbpfunit};
    \legend{rBPF, Femto-Containers, CertFC};
  \end{axis}}
  \end{tikzpicture}
  }
  \caption{Time per instructions on the Cortex-M4 platform}
  \label{fig:instructions2}
\end{figure}
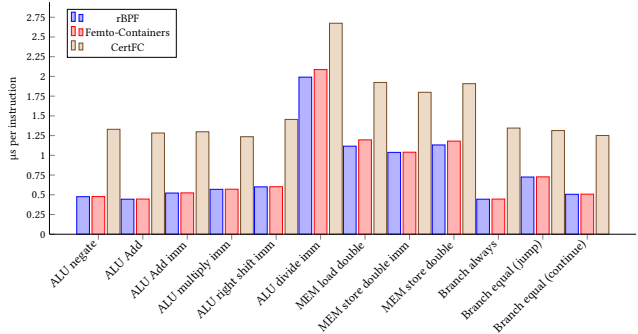

\subsection{Experiments with a Single Container}

In this section we show the execution times of a number of Femto-Container applications.
This shows the applicability of Femto-Containers in the different scenarios.
we show the execution times in \autoref{fig:application_exec}.

The first example executes a Fletcher32 checksum over a data string of \SI{360}{\byte}.
It shows the time it takes for relative heavy processing within the Femto-Containers VM.
Depending on the platform and the speed of the microcontroller it takes between \SI{1.3}{\milli\second} and \SI{2.2}{\milli\second}.
For Femto-Containers the duration of this application is long.

The second example shown is the thread counter example previously shown in \autoref{lst:thread_log}.
In normal operation it is inserted in the thread switch hook provided by the operating system, a hot path in the OS.
As shown in the figure, adding this would increase the duration of a thread switch in the operating system by \SI{10}{\micro\second} to \SI{27}{\micro\second}.
The impact on the operating system would not be negligible, but also not problematic during normal operation.

The last example shows the duration of the second stage of the networked sensor code example\cite{femto-fetch}.
It depends heavily on system calls for formatting of the CoAP response, but still contains some processing inside the VM.
It can be considered a representative example for business logic on the device.
This example takes between \SI{23}{\micro\second} and \SI{72}{\micro\second}.
For business logic programmed outside of the hot code path of the operating system itself, the overhead caused here by the VM is rather acceptable and doesn't impact the performance of the overall system.

\pgfplotstableread[col sep=comma,]{Measurements/fletcher.csv}\measurefletcher
\pgfplotstableread[col sep=comma,]{Measurements/threadlog.csv}\measurethreadlog
\pgfplotstableread[col sep=comma,]{Measurements/gcoapfmt.csv}\measuregcoapfmt
\begin{figure*}[ht]
\begin{subfigure}{.66\columnwidth}
\begin{tikzpicture}
  {\centering
  \begin{axis}[
        ybar, axis on top,
        axis x line*=bottom,
        axis y line*=left,
        height=5cm, width=\textwidth,
        enlarge x limits=true,
        ylabel={\si{\micro\second} per execution},
        ytick distance=500,
        ybar interval=0.7,
        ymin=0,
        xtick=data,
        xticklabel interval boundaries,
        xticklabels={Cortex-M4, ESP32, RISC-V},
        xticklabel style={font=\normalsize, rotate=90, anchor=east},
        legend style={
            anchor=north west,
            legend pos=north west,
        },
    ]
    \addplot table [x expr=\coordindex, y index=1]{\measurefletcher};
  \end{axis}}
  \end{tikzpicture}
  \caption{Fletcher32 checksumming algorithm application.}
\end{subfigure}
\begin{subfigure}{.66\columnwidth}
\begin{tikzpicture}
  {\centering
  \begin{axis}[
        ybar, axis on top,
        axis x line*=bottom,
        axis y line*=left,
        height=5cm, width=\textwidth,
        enlarge x limits=true,
        ylabel={\si{\micro\second} per execution},
        ytick distance=5,
        ybar interval=0.7,
        ymin=0,
        xtick=data,
        xticklabels={Cortex-M4, ESP32, RISC-V},
        xticklabel style={rotate=90, anchor=east},
        legend style={
            anchor=north west,
            legend pos=north west,
        },
    ]
    \addplot table [x expr=\coordindex, y index=1]{\measurethreadlog};
  \end{axis}}
  \end{tikzpicture}
  \caption{Thread log example application.}
\end{subfigure}
\begin{subfigure}{.66\columnwidth}
\begin{tikzpicture}
  {\centering
  \begin{axis}[
        ybar, axis on top,
        axis x line*=bottom,
        axis y line*=left,
        height=5cm, width=\textwidth,
        enlarge x limits=true,
        ylabel={\si{\micro\second} per execution},
        ytick distance=10,
        ymin=0,
        ybar interval=0.7,
        xtick=data,
        xticklabels={Cortex-M4, ESP32, RISC-V},
        xticklabel style={rotate=90, anchor=east},
        legend style={
            anchor=north west,
            legend pos=north west,
        },
    ]
    \addplot table [x expr=\coordindex, y index=1]{\measuregcoapfmt};
  \end{axis}}
  \end{tikzpicture}
  \caption{CoAP response formatter application.}
\end{subfigure}
  \caption{Execution duration of different examples running on Femto-Containers.}
  \label{fig:application_exec}
\end{figure*}

\subsection{Femto-Containers with Multiple Instances}

Femto-Containers are optimized to run multiple containers on a single system in parallel.
All state of an instance is kept local to the instance.
Each new instance added takes \SI{624}{\byte} of RAM to run, including the virtual machine stack.
The other requirement is that the microcontroller must have a large enough storage for the all the application images.

We now measure the memory required to concurrently host multiple containers from multiple tenants on the same microcontroller, from the examples we described in \autoref{sec:femto-container-examples}.
As shown previously in \autoref{tbl:femtosize}, the minimal default memory footprint used by a Femto-Container amounts to \SI{624}{\byte}, which is for storing the VM stack,  housekeeping structs and information about memory regions.
Furthermore, the key-value stores are also in RAM. In this case the total RAM used by the key value stores (and houskeeping) for different tenants was \SI{340}{\byte}.
Hence, the required RAM memory we measured so as to run the example with 3 containers and 2 tenants is \SI{3.2}{\kibi\byte}.
Beyond these examples, if we consider more containers hosting larger applications (\emph{e.g.} $\approx$\SI{2000}{\byte}) an Arm Cortex-M4 microcontroller with \SI{256}{\kibi\byte} RAM, the density of containers achievable would be of $\approx$100 instances, next to running the OS.

Different instances do not have access to each others resources by default.
They are fully isolated and do not have access to each others memory, isolated by the memory protection mechanism.
One way provided to communicate between the instances is the shared key-value store.

Multiple containers can be attached to the same launchpad hook inside the operating system.
It depends on the hook how the return value from each instance is processed further.
This allows for multiple tenants attaching to the same hook and process similar events.

\subsection{Overhead Added by Hooks}

One key question is how performance is affected by pre-provisioning launchpads (hooks) in the RTOS firmware.
We measure in \autoref{tbl:hook_comparison} the overhead caused by adding a hook to the RTOS workflow.
This overhead amounts to $\approx$100 clock ticks on all the hardware we tested.
Compared to the number of cycles needed for an average task in the operating system, this impact is low.
Furthermore, this overhead is less than 10\% of the number of cycles needed to execute the logic hosted in a Femto-Container.
From this observation, we can conclude that, even if this hook is on a very hot code path (as for the Thread Counter example) the performance loss is tolerable.
Conversely, the perspective of adding many hooks sprinkled in many places in the RTOS firmware is realistic without incurring significant performance loss.
% application  indeed relative low on the overal system performance.
%Also compared to the overhead of a Femto-Container application, the overhead of an empty hook is low, less than 10\% of the full execution with the thread switch example Femto-Container attached.

\begin{table}[ht]
  \centering
	\begin{tabular}{lrr}
		\toprule
		& Empty Hook & Hook with Application \\
		\midrule
		Cortex-M4 & 109 & 1750 \\
		ESP32 & 83 & 1163 \\
		RISC-V & 106 & 754 \\
		\bottomrule
  \end{tabular}
  \vspace{0.5em}
  \caption{Hook overhead in clock ticks for the thread switch example}
  \label{tbl:hook_comparison}
  \vspace{-1.5em}
\end{table}

\iffalse

\begin{figure}[t]
  \centering
  \begin{tikzpicture}
  \begin{axis}[
      domain=0:80,
      samples=80,
      smooth,
      no markers,
      ylabel = RAM footprint (KiB),
      xlabel = Number of Femto-Containers,
      legend pos=north west,
      scaled ticks=false,
    ]
    \addplot{((2000 + 660) * x)/1024};
    \addlegendentry{\SI{2000}{\byte} applications}
    \addplot{((500 + 660) * x)/1024};
    \addlegendentry{\SI{500}{\byte} applications}
    \addplot{((100 + 660) * x)/1024};
    \addlegendentry{\SI{100}{\byte} applications}
  \end{axis}
\end{tikzpicture}

  \caption{Density of Femto-Container instances.}
  \label{fig:femto-scaling}
\end{figure}

\fi

\section{Discussion}

%\paragraph{Processing overhead}
\paragraph{Virtualization \emph{vs} Power-Efficiency}
Inherently, virtualization causes some execution overhead, due to interpretation of the code.
Thus Femto-Containers increase power consumption for functionality executed within the \ac{VM}, compared to native code execution.
%This is an inherent limitation of the \ac{VM} technique.
However, this drawback is mitigated by several other factors.
First, the absolute power consumption overhead may be neglible, \emph{e.g.} if the hosted logic is not performing long-lasting, heavy-duty tasks.
Second, network transfer costs, power consumption and downtime are saved if software updates modify a Femto-Container instead of the full firmware.

\paragraph{Controlling Tenant Priviledges}
Controlling and granting access to specific RTOS resources to different containers or tenants is a complex challenge.
Our design includes a basic permission system based on preprovisionned hooks, system calls, and simple contracts between the hosting engine (on behalf of the OS) and a given container.
Basically: the OS restricts the set of priviledges that can be granted, the container specifies the set of priviledges it requires, and the hosting engine grants the intersection of these sets.
One limitation of our current simplified design is that there is only one fixed set of priviledges possible per hook.
In case 2 tenants have different priviledges, a second hook must be made available.
Additional mechanisms would be required to overcome this limiation and/or to enable dynamic priviledge levels.

\paragraph{Install Time \emph{vs} Execution Time}
As mentioned before, one limitation due to virtualization is the inherent slump in execution speed, compared to native code exection.
One way to remove this overhead is to transpile the portable eBPF bytecode into native instruction code.
This could be done in a single pass to convert the whole application into native instructions in an installation step.
This can result into a speed-up at the cost of extra install-time overhead.
To avoid the issues describe before on complicating the run-time security checks, this compilation into native code has to be done at run-time by the device deploying the code.

\paragraph{Fixed- \emph{vs} Variable-length Instructions}
Originally, eBPF scripts are optimized for fast execution on 64-bit platforms.
Compared to other virtual machines such as Wasm, the resulting bytecode is relative large.
In fact, most of the instructions have bit fields that are fixed at zero.
A possible way to reduce the size of these scripts is to compress the instructions into a variable size instruction set, removing these fields from the instructions where possible.
This would create a variable length instruction set based on the eBPF set.
For example the immediate field is not used with half of the instructions and would reduce the instructions to 32 bits in size when removed.

\section{Conclusion}

In this paper we have introduced Femto-Containers, a new middleware runtime architecture we designed, which enables %Function-as-a-Service (FaaS)
FaaS capabilities embedded on heterogeneous low-power IoT hardware.
Using Femto-Containers, authorized (3rd party) maintainers of IoT software can deploy and manage via the network mutually isolated software modules embedded on a microcontroller-based device.
We provided an open source implementation of the Femto-Container runtime, which uses the eBPF instruction set ported to microcontrollers, as well as integration in a common low-power IoT operating system (RIOT). 
We formally verified a fault-isolation guarantee which ensures that RIOT is shielded from arbitrary logic loaded and executed in a Femto-Container -- and such, without requiring any specific hardware-based memory isolation mechanism.
We then demonstrated experimentally the performance of the Femto-Container runtime on the most common 32-bit microcontroller architectures: Arm Cortex-M, RISC-V, ESP32.
We show that Femto-Containers significantly improve state of the art, by providing FaaS-like capabilities with strong security guarantees on such microcontrollers,
while requiring negligible Flash and RAM memory overhead (less than 10\%) compared to native execution.

%Femto-Containers improve state-of-the-art containerization, virtualization and eBPF use on IoT microcontrollers,by increasing security, isolation and execution speed. In effect, Femto-Containers enables hosting (tens of) applications executing concurrently, and multiple tenants, on a single low-power IoT device.

\iffalse
In this paper, we first compared and evaluated existing virtual machine and scripting implementations.
Based on these results we show that the existing eBPF instruction set shows a promising tradeof between code size, execution speed and application size.
We demonstrate the Femto-Containers virtual machine, an eBPF-based virtual machine hosted in RIOT.
We compare the performance against rBPF, the existing eBPF-based virtual machine for embedded systems.
We show that in terms of execution speed Femto-Containers performs better, while offering more security and isolation features, than rBPF.
We also show that the hosted applications on Femto-Containers are lightweight enough to run multiple in parallel on small embedded devices when considering small to medium software modules.
\fi

%-------------------------------------------------------------------------------
%\section*{Availability}
%-------------------------------------------------------------------------------

%The implementation of Femto-Containers is open source \cite{femto-github}.

%-------------------------------------------------------------------------------
%\section*{Appendix A: Benchmark Hardware \& Software Configuration}
\section*{Appendix A: Benchmark Configuration}
{\bf Hardware --} We carry out our measurements on popular, commercial, off-the-shelf IoT hardware, representative of the landscape of the modern 32-bit microcontroller architectures that are available. More precisely, we build and run the code on the following boards, all configured to run at \SI{64}{\mega\hertz}:
\begin{itemize}
    \item {\bf Arm Cortex-M:} a Nordic nRF52840 Development Kit, using an Arm Cortex-M4 microcontroller
        with \SI{256}{\kibi\byte} RAM,
        \SI{1}{\mebi\byte} Flash,
        and a \SI{2.4}{\giga\hertz} radio transceiver\\ (BLE/802.15.4),
        %compatible with both IEEE 802.15.4 and Bluetooth Low-Energy.
    \item {\bf ESP32:} a WROOM-32 board, using an Espressif ESP32 module %with the ESP32-D0WDQ6 chip on board),
        which provides two low-power
        Xtensa\textsuperscript{\tiny\textregistered} 32-bit LX6
        microprocessors with integrated Wi-Fi and Bluetooth,
        \SI{520}{\kibi\byte} RAM,
        \SI{448}{\kibi\byte} ROM
        and \SI{16}{\kilo\byte} RTC SRAM.
    \item {\bf RISC-V:} a Sipeed Longan Nano GD32VF103CBT6 Development Board,
        which provides a RISC-V 32-bit microcontroller
        with \SI{32}{\kibi\byte} RAM
        and \SI{128}{\kibi\byte} Flash.
\end{itemize}

Note that an open-access testbed such as IoT-Lab~\cite{iot-lab2015-ieee-iot-wf} also provides some of this hardware, for reproducibility.\ \\

\noindent {\bf Software --} In all benchmarks, the embedded software platform (OS) hosting the Femto-Containers is RIOT~\cite{baccelli2018riot}.
As base, we took RIOT Release 2022.01, configured to be IoT-ready. 
More precisely, we configured RIOT to provide standard low-power networking connectivity, leveraging the board's IEEE 802.15.4 radio chip and a ressource-efficient IPv6-compliant network stack (6LoWPAN, UDP, CoAP), as well as providing secure software update capability, enabling the update of Femto-Containers over the low-power network, in compliance with SUIT~\cite{suit-manifest} specifications (using CBOR, COSE, ed25519 signatures and SHA256 hashes as primitives).

\section*{Acknowledgements}
The research leading to these results partly received funding from the RIOT-fp project, and from the TinyPART project (within the MESRI-BMBF German-French cybersecurity program under grant agreements no ANR-20-CYAL-0005 and 16KIS1395K). The paper reflects only the authors’ views. MESRI and BMBF are not responsible for any use that may be made of the information it contains.



\bibliography{bibliography}
\bibliographystyle{unsrt}
%\bibliographystyle{ACM-Reference-Format}

\iffalse

\appendix

\section{Research Methods}

\subsection{Part One}

Lorem ipsum dolor sit amet, consectetur adipiscing elit. Morbi
malesuada, quam in pulvinar varius, metus nunc fermentum urna, id
sollicitudin purus odio sit amet enim. Aliquam ullamcorper eu ipsum
vel mollis. Curabitur quis dictum nisl. Phasellus vel semper risus, et
lacinia dolor. Integer ultricies commodo sem nec semper.

\subsection{Part Two}

Etiam commodo feugiat nisl pulvinar pellentesque. Etiam auctor sodales
ligula, non varius nibh pulvinar semper. Suspendisse nec lectus non
ipsum convallis congue hendrerit vitae sapien. Donec at laoreet
eros. Vivamus non purus placerat, scelerisque diam eu, cursus
ante. Etiam aliquam tortor auctor efficitur mattis.

\section{Online Resources}

Nam id fermentum dui. Suspendisse sagittis tortor a nulla mollis, in
pulvinar ex pretium. Sed interdum orci quis metus euismod, et sagittis
enim maximus. Vestibulum gravida massa ut felis suscipit
congue. Quisque mattis elit a risus ultrices commodo venenatis eget
dui. Etiam sagittis eleifend elementum.

Nam interdum magna at lectus dignissim, ac dignissim lorem
rhoncus. Maecenas eu arcu ac neque placerat aliquam. Nunc pulvinar
massa et mattis lacinia.

\fi

\end{document}